\documentclass[12pt]{iopart}

\usepackage{iopams}  
 \usepackage{graphicx}
 
\begin{document}
 \def\Journal#1#2#3#4{#1 {\textit{#2}} {\bf #3} {#4} } 
 
% Some useful journal names 
\def\BiJ{ Biophys. J.}                 
\def\Bios{ Biosensors and Bioelectronics} 
\def\LNC{ Lett. Nuovo Cimento} 
\def\JCP{ J. Chem. Phys.} 
\def\JAP{ J. Appl. Phys.}
\def\JACS{ J. Am. Chem. Soc.} 
\def\JMB{ J. Mol. Biol.} 
\def\CMP{ Comm. Math. Phys.} 
\def\LMP{ Lett. Math. Phys.} 
\def\NLE{{ Nature Lett.}} 
\def\NPB{{ Nucl. Phys.} B} 
\def\PLA{{ Phys. Lett.}  A} 
\def\PLB{{ Phys. Lett.}  B} 
\def\PRL{ Phys. Rev. Lett.} 
\def\PRA{{ Phys. Rev.} A} 
\def\PRE{{ Phys. Rev.} E} 
\def\PRB{{ Phys. Rev.} B} 
\def\PNAS{{ Proc. Natl. Acad. Sci.}}
\def\EPL{{Europhys. Lett.} } 
\def\PD{{ Physica} D} 
\def\ZPC{{ Z. Phys.} C} 
\def\RMP{ Rev. Mod. Phys.} 
\def\EPJD{{ Eur. Phys. J.} D} 
\def\SAB{ Sens. Act. B} 
\title[Photoreceptors for a light biotransducer]{Photoreceptors for a light biotransducer: a comparative study of the electrical responses of  two \textit{(type-1)} opsins}
% Force line breaks with \\

\author{E Alfinito$^{1,3}$, J Pousset$^2$, L Reggiani$^{2,3}$ and K Lee$^4$}
\address{$^1$ Dipartimento di Ingegneria dell'Innovazione, Universit\`a del
Salento, via Monteroni, I-73100 Lecce-Italy (EU)}
\address{$^2$ Dipartimento di Matematica e Fisica "Ennio De Giorgi", Universit\`a del
Salento, via Monteroni, I-73100 Lecce-Italy (EU)}
\address{$^3$ CNISM - Consorzio Nazionale Interuniversitario per le Scienze Fisiche della Materia, via della Vasca Navale, 84, 00146  Roma-Italy (EU)}
\address{$^4$ School of Chemical and Biological Engineering, Institute of Bioengineering, Seoul National University, Seoul 151-742, Republic of Korea
}
\ead{eleonora.alfinito@unisalento.it}
\ead{jeremy.pousset@unisalento.it}
\ead{lino.reggiani@unisalento.it}
\ead{klee@sogang.ac.kr}
%

% \textbackslash\textbackslash
%

\begin{abstract}
The increasing interest in photoactivated proteins as natural replacement of standard inorganic materials in photocells drives to the compared analysis of bacteriorhodopsin and proteorhodopsin, two widely diffused proteins belonging to the family of \textit{type-1} opsins. 
These proteins share similar behaviours but exhibit relevant differences in the sequential chain of the amino acids constituting their tertiary structure.
The use of an impedance network analogue to model the protein main features provides a microscopic interpretation of a set of experiments on their photoconductance properties. 
In particular, this model links the protein electrical responses to the tertiary structure and to the interactions among neighbouring amino acids.
The same model is also used to predict the small-signal response in terms of the Nyquist plot. 
Interesting enough,  these rhodopsins are found to behave like a wide gap semiconductor with intrinsic conductivities of the order of $10^{-7}$ S/cm.
\end{abstract}
%87.15.Zg       Phase transitions (biomolecules)

%\maketitle

\submitto{\NT}
\section{Introduction}
The growing demand of eco-friendly fuels is orienting the research toward
the production of photovoltaic devices based on new kinds of active matter.
In  particular,  organic and biological materials should guarantee better
performances when compared with standard solid-state materials, like semiconductors, oxides, etc. 
Therefore, much interest  arose on  photoactive proteins, i.e. biomaterials able to convert, \textit{in vivo}, visible light in energy useful for the cell survival. 
Among these proteins, the best known is  bacteriorhodopsin (bR), which uses green light to pump protons outside the cellular membrane \cite{oerst}.
This protein,  with some lipids, constitutes the so-called purple membrane of the Archean microorganism \textit{Halobacterium salinarum}.
Purple membrane appears as a film of about 5 nm thickness, with the proteins close packed in hexagons.
\par
Another photoactivated protein is recently receiving constant monitoring: the proteorhodopsin (pR), found in an uncultivated marine bacterium (SAR-86 group), which shows features more complex than bR and of wider interest in technology and ecology \cite{Beja,Beja1,Lorinczi,Kralj,Riesenfeld,Johnson}.
For example, different kinds of pRs were found in marine bacteria, usually adapted to different wavelengths, at different ocean depths \cite{Beja, Beja1,Riesenfeld}.
Furthermore, it seems quite easy to produce protein mutants, able to react to different wavelengths and also to be engineered for producing light flashes \cite{Kralj}.
Among pR peculiarities we also recall the possibility of inward or outward proton pumping with respect to the value of the environmental pH concentration \cite{Lorinczi}.
Moreover, a substantial improvement of current generation can be observed during illumination in electrochemical chambers containing bacteria in which pR was expressed \cite{Johnson}. 
\par
Finally, pR displays promising technological applications: it is abundant in nature, easy to express in different proteins \cite{Lorinczi,Kralj,Riesenfeld,Johnson}, it works with different wavelengths (different mutants) and thus  it is a perfect dye for optoelectronic devices. 
In spite of these large potentialities, till  now a few attention has been  devoted to identify pR  tertiary (3D) structure which, in a first attempt,  is assumed similar to that of bR, with seven transmembrane helices \cite{Reckel}. 
On the other hand, the knowledge of the protein topology is a relevant information for
investigating the protein features, like, for example, the differences among mutants, the
electrical properties, etc.  \cite{Nano,Jap,EPL,PRE,PRE1}. 
\par
Light-activated proteins like bR and pR are conjugated with a retinal molecule located among the seven trans-membrane helices of these proteins. 
The retinal isomerizes going from the \textit{all-trans }to the \textit{13-cis} configuration when photons of the visible-light spectrum hit it. 
As a consequence of this change, a cycle of transformations involves the protein as a whole, driving it from the not-active (native) state to the active state and, in turn, back to the native state (conformational change). 
During this cycle, a proton is first released and then re-collected. 
The structure change is expected to produce some modifications in the electrical properties of the protein and this aspect is crucial for any possible use of this (or similar) protein in light converter devices \cite{Nano,Jap,EPL,PRE,PRE1}.    
\par
To explore these features, some studies were performed on bR, in particular on its electrical response when sandwiched between metallic contacts in a metal-insulator-metal (MIM) structure \cite{Jin,Jin1,Ron,Gomila}. 
Bacteriorhodopsin exhibits a conductivity close to that of wide-gap semiconductors, and is robust against thermal, chemical and photochemical degradation \cite{Jin,Jin1}. 
Junctions prepared with  monolayers of bR including their lipids and contacted with Au electrodes,
show super-linear current-voltage (I-V) characteristics both in dark (or blue light) and in  the presence of green light \cite{Jin,Jin1,Ron,Gomila}. 
Green light produces a higher current with respect to dark conditions, as a consequence of the modifications of the chromophore-opsin complex \cite{Jin,
Jin1}. 
The super-linear I-V characteristics point toward a tunnelling charge-transfer mechanism \cite{Jin,Jin1,Gomila,EPL,PRE1}.
\par 
First measurements of I-V characteristic in pR were recently reported in Ref.~\cite{Lee}. 
The experiment was performed on a thin-film device with a contact distance of 50 $\mu m$, just about $10^{4}$  times longer than that used in bR nanolayers \cite{Jin,Jin1,Gomila}. Also in this case, like as in bR, a significant current
response was detected both in dark and in light.
The lack of the knowledge of the activated 3D structure of pR and the different procedure
of the experiments, make
%Furthermore, the activated tertiary structure of pR is at present not available %in the literature.
%Therefore,  
a direct comparison between data obtained in pR and in bR not easy to be carried out, especially at a microscopic level. 
%Anyway, 
%in both these proteins, the presence of a significant photocurrent, i.e. %the difference between the light and dark current response, makes 
%these
%proteins appear of great interest for optoelectronic applications.  
On the other hand, phenomenological approaches provide interpretations of some I-V characteristics in terms of a direct MIM tunneling expression \cite{Gomila,Reed}. 
However, these approaches are not able to describe the protein electrical modification induced by different conformations and or conformational changes.
\par
In this paper we investigate the small-signal response and the high-field electrical properties of pR in connection with its 3D structure.
Because of the strict analogies between pR and bR, we  carry out a comparative analysis of these proteins, thus providing expectations to be confirmed by future experiments. 
The investigation is performed by completing existing experiments \cite{Lee} and using a theoretical approach,  the impedance network protein analogue (INPA), already validated on similar biological systems\cite{EPL,PRE,PRE1}. 
Finally we show how all the experimental results can be interpreted within a unified theoretical framework.
\par
The content of the paper  is organized as follows. 
Section 2 reports new experimental results on pR which complement those in Ref.~\cite{Lee}. and provides their interpretation in terms of phenomenological laws. 
Section 3 recalls the INPA microscopic approach and reports the numerical results to be compared with experiments, starting from the tertiary structure of the protein. 
Major conclusions are drawn in Sec. 4.  
Finally, the Appendix details the main structural and electrical differences among the bR and pR protein models that are used in numerical calculations.  
\section{Experiments}
Bacteriorhodopsin and proteorhodopsin are trans-membrane proteins characterized by seven helices that cross the cell lipid-membrane; in this way, they allow the internal side of the cell to communicate with the external environment.
Their  similarities with the G-protein coupled receptors and their  easy-to-manipulate properties make these proteins useful templates to infer the structure of different trans-membrane proteins, for example, by homology modeling.
The first stage of the sensing is the photon absorption, then followed by  a conformational change of the protein tertiary structure.
Accordingly, this change was proven to be sufficient to modify the electrical properties of the single protein \cite{Jin,Jin1,PRE,PRE1,EPL}.
In particular,  experiments performed on patches of bR purple membrane MIM structures revealed a significant electrical current in dark and a noticeable photocurrent,  i.e. the difference between the light and dark current response, in the presence of green  light.
This property is not a prerogative of bR, as a recent set of  measurements performed on a MIM thin film of pR have shown \cite{Lee}. 
In this latter experiment, the channel width of the film active region is of about $50 \ \mu$m, and it is  sandwiched between two gold electrodes evaporated on a glass substrate with thickness of about 200 nm  and length of about 3 mm.
Details on the experiment are reported in Ref.~\cite{Lee}; here we simply notice the main differences with respect to previous experiments on bR: (i) the distance between electrodes and, (ii)  the sample preparation. 
Indeed, in pR experiments the distance between electrode is for about a factor of  $10^{4}$  greater  than that in bR experiments.
Furthermore, pR  measurements are performed on proteins diluted in a buffer solution and then dried; in the case of bR they were performed directly on the membrane film.
Nevertheless, also in pR  a dark current and a photocurrent are detected, thus confirming that also \textit{in vitro} a photoactivated protein can work like a  light-current converter.  
\par
In the following we report  new results on the electrical current in pR, both in dark and in light, obtained  with the experimental set up described in Ref.~\cite{Lee}.
These data complement those already  reported and admit for a direct  interpretation in terms of phenomenological laws.
\begin{figure}[htbp]
        \centering
                \includegraphics[width=0.35\textwidth,angle=-90]{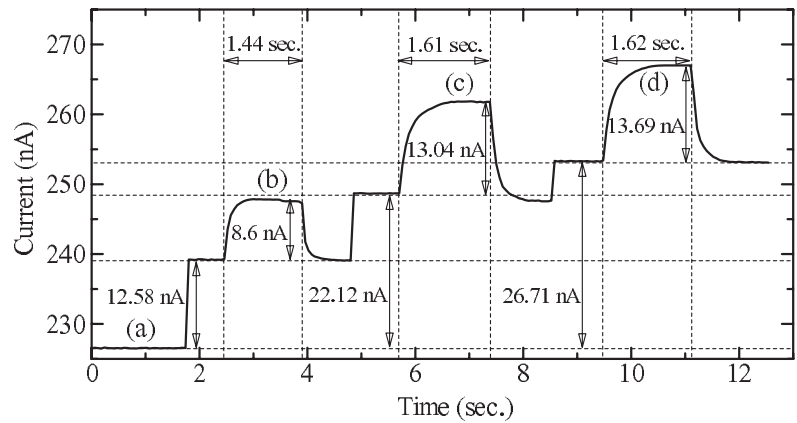}
        \caption{Real time measured photocurrent for (a) buffer alone, (b) to (d) buffer and  pR  with concentration of:  1 OD, 5 OD, and 10 OD under
150  mW/cm$^2$ incident light intensity at 625 nm wavelength. The voltage applied to the MIM system was 120 V. The vertical dashed lines are marks to show the exposure times in every case.
}
\label{fig:ivst}
\end{figure}
%fig1
%
\par
\Fref{fig:ivst} shows the real-time measured photocurrent for (a) buffer alone and, (b) to (d) buffer plus pR with concentrations of (b) 1 OD, (c) 5 OD, and (d) 10 OD for 150 mW/cm$^{2}$ incident light intensity at 625 nm wavelength and 120 V fixed voltage.  
The observed current level is constant (227 nA) for buffer alone, but increases when adding pR with different concentrations and also when, for a given pR concentration, the sample is illuminated by an incident light. 
The current increase over the buffer level in dark conditions is, respectively, of 12.58 nA, 22.12 nA, and 26.71 nA.
From \Fref{fig:ivst} the estimated photocurrents are 8.6 nA, 13.04 nA, and 13.69 nA, respectively for 1 OD, 5 OD, and 10 OD. 
\par
\Fref{fig:ivst} proves the existence of a contribution to the dark current of the film due to pR, which adds to that associated with the buffer only, and of a net photocurrent in the presence of a visible light, also due to  pR.  
Both these dark and light current contributions increase at increasing pR concentration.
As a further consideration, we notice that the photocurrent in pR is of about 68 \% higher than the dark current for a concentration of 1 OD, and of  about 51 \% higher for a concentration of 10 OD.  
Such a decrease  of the photocurrent at increasing pR concentration should be explained with a stratified growth  of the pR proteins on the surface of the film, which partially shields the absorption of  impinging photons.
\par
The information collected in \Fref{fig:ivst} can be described within the Langmuir adsorption law \cite{langmuir16} that predicts a saturation after an initial linear increase of the current as:
\begin{equation}
I=G_{\rm {tot}} V 
\label{eq:lang}
\end{equation}
with
\begin{equation}
G_{\rm {tot}}(C)= G_{\rm {sub}} + \frac{C}{1+C} G_{\rm {pRsat}} 
\label{eq:lang1} 
\end{equation}
where $C$ is the pR concentration in units of OD. 
\par
\begin{figure}[htbp]
        \centering
                \includegraphics[width=0.6\textwidth]{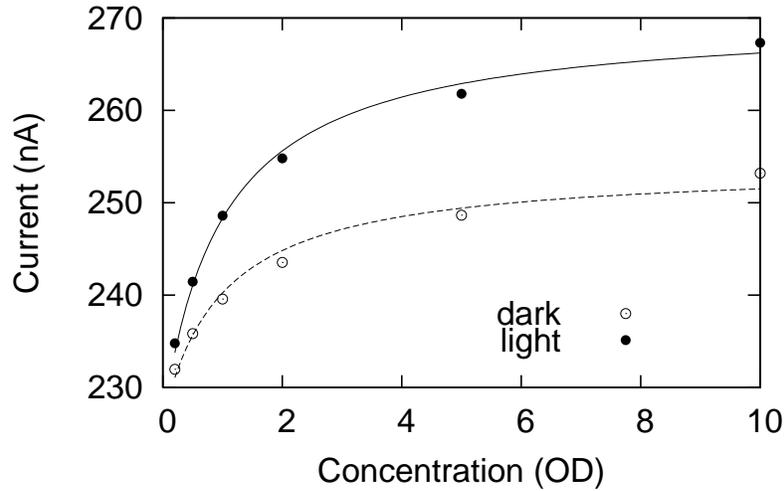}
        \caption{Current as function of pR concentration at 120 V  applied voltage. Open (full) circles refer to dark (illuminated) conditions and dashed (continuous) curves to a fit of 
experiments obtained with the expression reported in text.
}
\label{experiments_1}
\end{figure}
%
%fig2
%
\Eref{eq:lang} is used in \Fref{experiments_1} to fit currents in dark (open circles) and in light 
(full circles) as a function of the pR concentration. 
In particular, the best fit parameters it are 
$G_{\rm {sub}}=1.89$ nS, and $G_{\rm {pRsat}} =0.229$ nS  for the dark and  $G_{\rm {pRsat}}=0.363$ nS for the illuminated conditions, with an intensity
 of 150 mW/cm$^2$. 
\begin{figure}[htbp]
        \centering
                \includegraphics[width=0.6\textwidth]{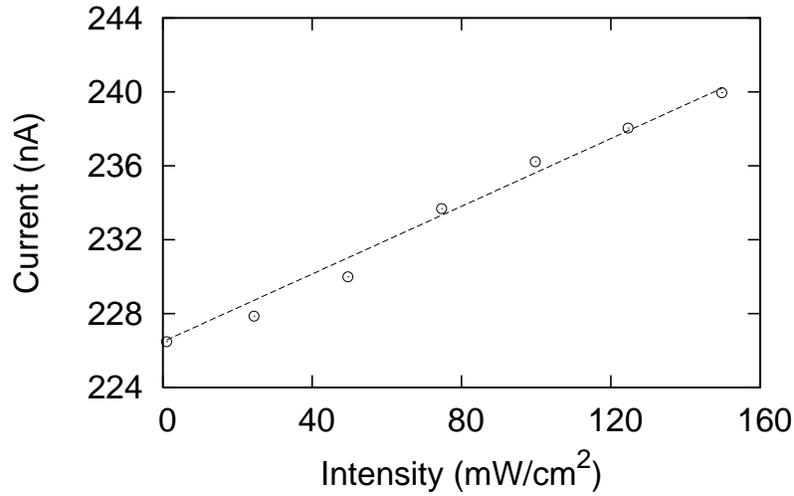}
        \caption{Total current flowing through the device  as a function of light intensity for an applied voltage of 120 V and a pR concentration of 10 OD.}
        \label{experiments2}
\end{figure}
%
%fig3
%
\par
The experiments exhibit a quasi-linear increase of photo-current at increasing light intensity. This behaviour is thus fitted by the expression:
\begin{equation}
I_{\rm {tot}}(P)=0.0916 \times P + I(0) 
\end{equation}
with $I(0) = 227$  nA, as reported in \Fref{experiments2}. 
Notice that in the examined range of radiation power, no saturation effect is observed.
\par
\begin{figure}[htbp]
        \centering
                \includegraphics[width=0.35\textwidth,angle=-90]{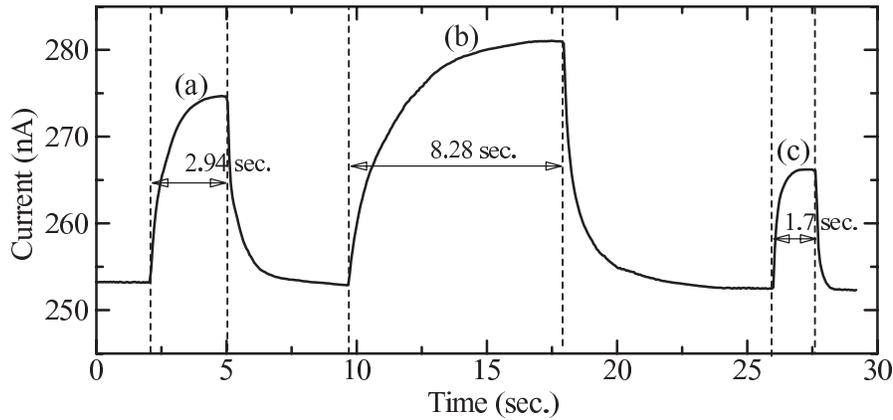}
        \caption{Real time measured photocurrent of 10 OD pR at incident light wavelengths of: (a) 472 nm, (b) 532 nm, and (c) 625 nm. The incident light intensity and applied voltage are 150 mW/cm$^2$ and 120 V, respectively.}
        \label{Lee2_red}
\end{figure}
%fig4
%
To test the pR sensitivity to different light wavelengths, \Fref{Lee2_red} reports the time diagram of the output photocurrent of 10 OD pR at the incident light, respectively of: (a) 472 nm (blue), (b) 532 nm (green) and (c) 625 nm (red). 
The intensity of the incident light is of 150 mW/cm$^2$, with an applied bias of 120 V.
In \Fref{Lee2_red} three independent but similar measurements are combined, for comparison with each other. 
In terms of amplitude, the photocurrent seems to be proportional to the absorption coefficient, as expected for efficient light conversion.
Furthermore, it can be noticed that the response time to the light excitation is related to the photocurrent amplitude.
Accordingly, the photo-generation phenomena is slowest in the green range (8.28 s), faster in the blue range (2.94 s), and finally the fastest in the red range (1.7 s), as shown in \Fref{Lee2_red}. 
This reminds the behaviour of  conventional photo-conductors where the gain-bandwidth product is constant \cite{rose78}: the photocurrent signal then becomes higher as the time response becomes longer.
\par
Finally, \Fref{Lee2_red} illustrates the high selectivity of pR samples to green light (532 nm). 
The blue-light response of pR samples should be associated with the coexistence of proteins in the activated M-state (maximum absorption at 410 nm) and proteins in the native state, as suggested by recent models of protein dynamics \cite{Deupi,RCSA}.
\section{Microscopic modeling}
\label{Modeling}
The role played by the 3D protein structure in the charge transport mechanism
is emphasized in several experimental papers \cite{Jin, Jin1, Ron}. These
papers clearly show that, in the same experimental conditions, different
proteins \cite{Ron} or different protein conformations \cite{Jin, Jin1} give
different current-voltage characteristics. Therefore, an approach looking at the
microscopic structure seems quite appropriate to describe
the protein electrical properties.
\par
To provide a microscopic interpretation of experiments, we describe the single protein electrical properties as a function of the protein morphology.
To this purpose we make use of the INPA modeling  already detailed in previous papers \cite{Nano,Jap}. 
In short, each protein is associated with a topological network (graph), consisting of nodes and links. 
Once the tertiary structure of the protein is given by the protein data base (PDB) \cite{PDB} or similarly, the graph reproduces its main features.
To this purpose, the nodes are in correspondence with the amino acids, taken as single interacting centers, and their position coincides with the related $C_\alpha$ atoms. 
Each couple of  nodes is connected with a link when their distance is less than  an assigned interacting radius, $R_c$. 
In this way, the protein network analogue becomes a set of intercrossing spheres of radius  $R_c$. 
In principle, all the values of $R_c$, from zero to a few nanometers (the protein size) are possible. 
Actually,  the choice of the $R_c$ value  depends on several factors, like the kind of interaction under consideration \cite{Tirion,Juanico,Nano,Jap,PRE}, the role of environment/ligands \cite{RCSA} and, marginally, on the protein size. 
In the present case, the only interaction we take into account is that associated with the transfer of charge between neighborhood amino acids, and $R_c$ is taken in the range of values 
$5 \div 60$ \AA, as detailed in Appendix.
Following the indication of experiments \cite{Jin, Jin1,Ron}, we assume the charge transfer through the protein as the most important transport mechanism and adopt this model to interpret all available experiments.
%%%%%%%%%%%%%%%%%%%
Further implementation of the model to account for other channels of charge transfer, e.g. through the surrounding which contains ions and water molecules as well as between proteins, is not found  necessary at present, and thus it is left to future research when also more data should be available.   
\par
The graph turns into an impedance network when an elemental impedance is associated with each link. 
Following the experimental outcomes, here the elemental impedance between the $i$-th and $j$-th nodes, say, $Z_{i,j}$ is taken as \cite{Nano}:
\begin{equation}  
Z_{i,j}=\frac{l_{i,j}}{{\mathcal{A}}_{i,j}}    
\frac{1}{(\rho^{-1} + \rmi \epsilon_{i,j}\, \epsilon_0\omega)} \label{eq:1}
\end{equation} 
where ${\mathcal{A}}_{i,j}=\pi (R_{c}^2 -l_{i,j}^2/4)$, is the cross-sectional area between the  spheres centered on the $i,j$ nodes, respectively, $l_{i,j}$ is the distance between these centers, $\rho$ is the  resistivity, taken to be the same for any  amino acid, (here the microscopic mechanism responsible of charge transfer is not specified), $\rm{i}=\sqrt{-1}$  is the imaginary unit, $\epsilon_0$ is the vacuum permittivity and  $\omega$ is the circular frequency of an applied harmonic voltage.  
The relative dielectric constant pertaining to the couple of $i,j$ amino acids, $\epsilon_{i,j}$, is expressed in terms of the intrinsic  polarizability of the $i,j$ amino acids \cite{Nano}.
\par
By positioning the input and output electrical contacts on the first and last node, respectively, for a given applied bias the network is solved within a linear Kirchhoff's scheme and its global impedance spectrum, $Z(\omega)$, is calculated in the standard frequency range $ 0.1 \div 10^5$ Hz.  
By construction, besides the small-signal  dynamic response, this network produces a parameter dependent static I-V characteristic determined as 
\begin{equation}
V = Z(0) I
\end{equation}
To account for the super linear current at high applied voltages, a tunnelling mechanism of charge transfer is included by using a stochastic approach \cite{PRE,PRE1} and, in particular, by allowing the elemental impedance to change its value accordingly to an assigned probability transition function. In this case, each link is considered as a simple resistive element, and the resistivity $\rho(V)$ depends only on the potential drop $V_{i,j}$ across the $i-j$ couple of nodes:
\begin{equation}
\rho(V)=\left\{\begin{array}{lll}
\rho_{MAX}& \hspace{.5cm }& eV_{i,j} \leq  \Phi  \\ \\
 \rho_{MAX} (\frac{\Phi}{eV_{i,j}})+\rho_{min}(1- \frac{\Phi}{eV_{i,j}}) &\hspace{.5cm} & eV_{i,j} \ge  \Phi 
 \end{array}
  \right.
\label{eq:3}
\end{equation}
where $\rho_{MAX}$ is the resistivity value which should be used to fit the I-V characteristic at the lowest voltages,  $\rho_{min} \ll \rho_{MAX}$  plays the role of an extremely low series resistance, limiting the current at the highest voltages,  and 
$\Phi$  is the effective height of the tunneling barrier.
The transmission probability function is the following :
\begin{equation} 
\mathcal{P}_{i,j}= \left\{
\begin{array}{lll}
\exp\left[-
\frac{2l_{i,j}}{\hbar} \sqrt{ 2m\left(
\Phi-\frac{1}{2} eV_{i,j}\right)}\right] 
&\hspace {.5cm} & 
 eV_{i,j}  \leq \Phi \\ \\
 \exp \left[-
 \left(
 \frac{2l_{i,j}\sqrt{2m}}{\hbar}\right)
 \frac{\Phi}{eV_{i,j}}\sqrt{\frac12\Phi} \right] 
& \hspace{.5cm} & eV_{i,j} \ge \Phi  
 \end{array}
  \right.
  \label{eq:1}
\end{equation}
where  $m$  is the electron effective mass, here taken the same of the bare value. 

\par
The spatial distribution of the obtained network is found to be in general highly irregular. 
Accordingly, the network describes the protein topology only when the tree structure is dominant, i.e. when the value of $R_c$ is sufficiently small so that the links between nearest neighbouring predominate. 
We remark, that the tree structure might not  suffice to reproduce the  global functioning of the protein. 
Thus,  it is necessary to find an optimal value of $R_c$ which can reproduce both  the interacting map and the electrical properties of the protein \cite{Jap,Juanico}. 
From one hand, the $R_c$ value should be greater than the mean distance between contiguous $C_\alpha$ atoms, say 3.8 \AA, otherwise the network appears as not connected. 
From  another hand, for very large values of $R_c$, say 60 \AA, each node will link all the others and the network is no longer  descriptive of the protein topology.
Accordingly, it was  shown \cite{Nano,Jap} that the best agreement between structure and function  is obtained with values of $R_c$ in the range $5 \div 60$ \AA. 
\par
On the side of the protein structures, while for bR in the native and active state more and more refined models have been proposed, for pR there is only one certified set of data, taken within  NMR technique, describing the protein in its native state. 
This set, the 2L6X entry of the PDB \cite{PDB,Reckel}, consists of 20 different models  of pR,  in principle equally able to describe the protein electrical properties.
\par
In the following, the investigation of the protein modifications is performed by means of different tools, such as: (i) the topological analysis, (ii) the determination of the small-signal response and, (iii) the determination of the static I-V characteristics at arbitrary strength of the applied electric field.
The investigation is carried out through a  comparative analysis of pR and bR  properties.
\subsection{Topological analysis}
This section investigates the possibility to infer the main electrical differences among the models proposed by the NMR analysis. 
As a matter of fact, instead of a single structure, the NMR method produces a family of 3D protein structures (models). 
In the present case, some main dissimilarities seem to be due to the large differences in the determination of the position of the binding pocket and of some loop arrangements \cite{Reckel}.
\begin{figure}
\centering
\includegraphics[width=0.35\textwidth]{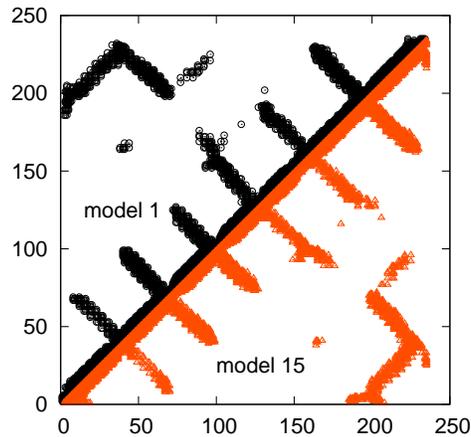}
\caption{Contact maps of the models 1 (on the left) and 15 (on the right) of a single protein of pR in its native state taken for $R_{c}$=12 \AA.}
\label{fig:contactmap_1_vs15}
\vskip1pc
\end{figure}
%fig5
For the 20 models of PDB entry 2L6X, the structure differences are analyzed by means of the contact maps which, for an assigned $R_{c}$ value, show the connected amino acids.
\par
Figure 5 reports the contact maps of models 1 and 15 of pR traced for $R_{c}=12$ \AA.
Without loss of generality, model 1 is taken as the reference structure.
Significant differences can be observed between model 1 and model 15, in particular in the central region of the protein (the same is found for other models, here not shown). 
On the other hand, when comparing pR with bR, as reported in Fig. 6, the similarities between these proteins clearly emerge and also the differences are found to be more significant around the central zone  of the protein.
\begin{figure}
\centering
\includegraphics[width=0.35\textwidth]{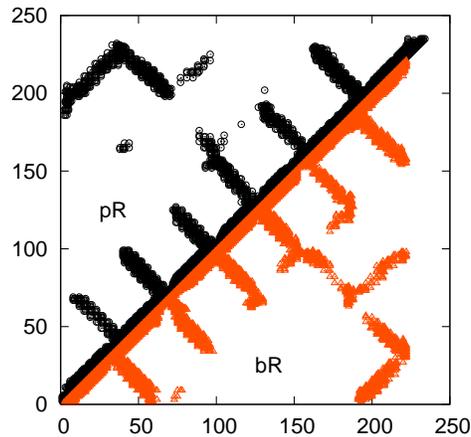}
\caption{Contact maps of the models 1 of a single protein of pR (on the left) and bR (on the right) in their native state taken for $R_{c}$=12 \AA.}
\label{fig:contact_map_rho_proteo_r12}
\vskip1pc
\end{figure}
%fig6
%
%
%\section{A comparative analysis of pR and bR electrical properties}
%
By construction, each of the 20 NMR models produces  different electrical responses when inserted in the INPA model. 
Therefore, it is useful to analyze these differences in order to state the boundaries of theoretical expectations on the basis of the present information.
\subsection{Small-signal response}
\begin{figure}\centering
        \begin{minipage}[c]{16pc}
        \includegraphics[width=0.9\textwidth]{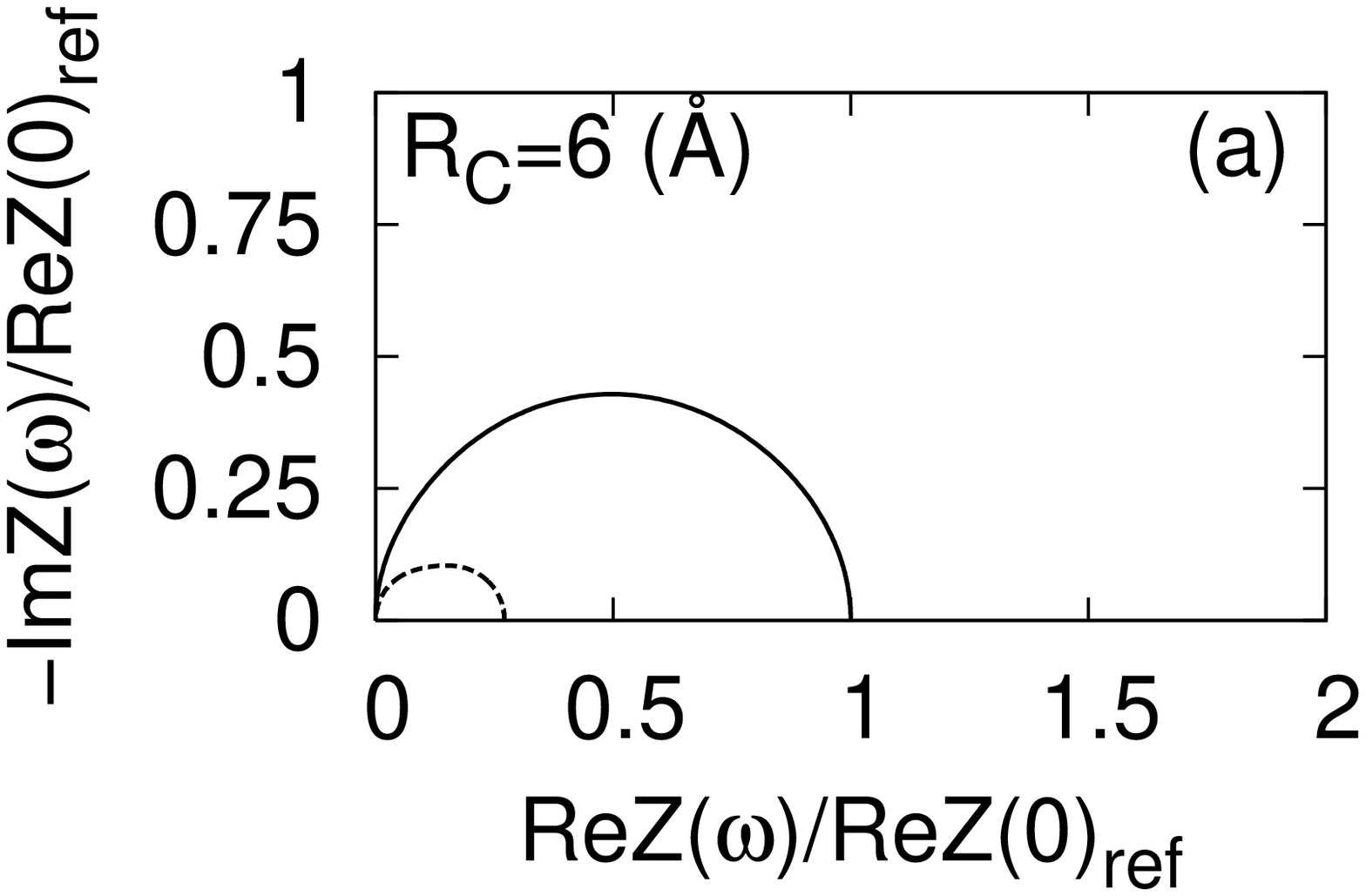} 
%\caption{minipage} 
\end{minipage} 
%\hspace{3cm}
\begin{minipage}[c]{18pc}
        \includegraphics[width=0.9\textwidth]{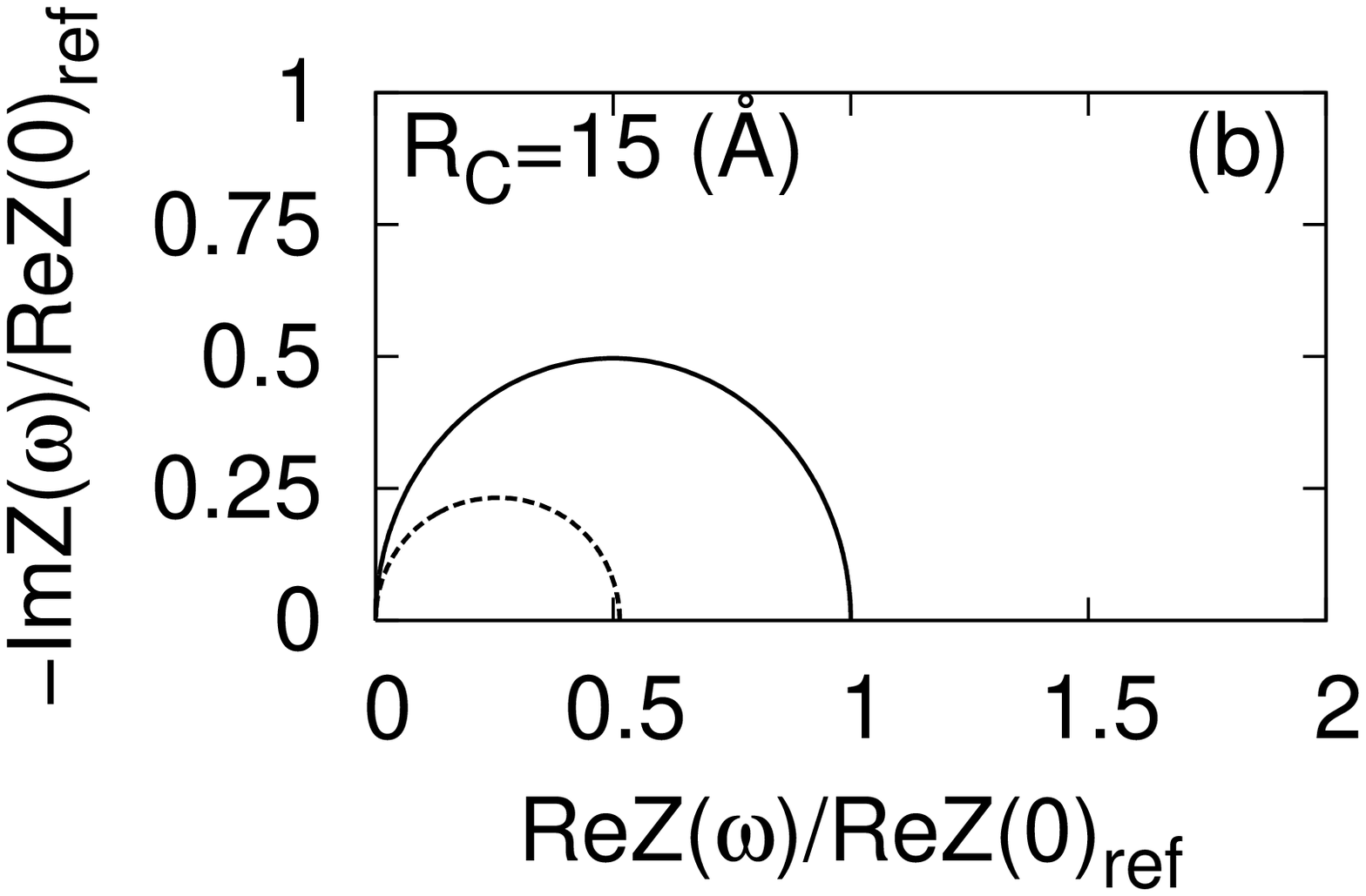} 
%\caption{minipage} 
\end{minipage} \\ 
%\hspace{2cm}
\vspace{1cm}
        \begin{minipage}[c]{18pc}
        \includegraphics[width=0.9\textwidth]{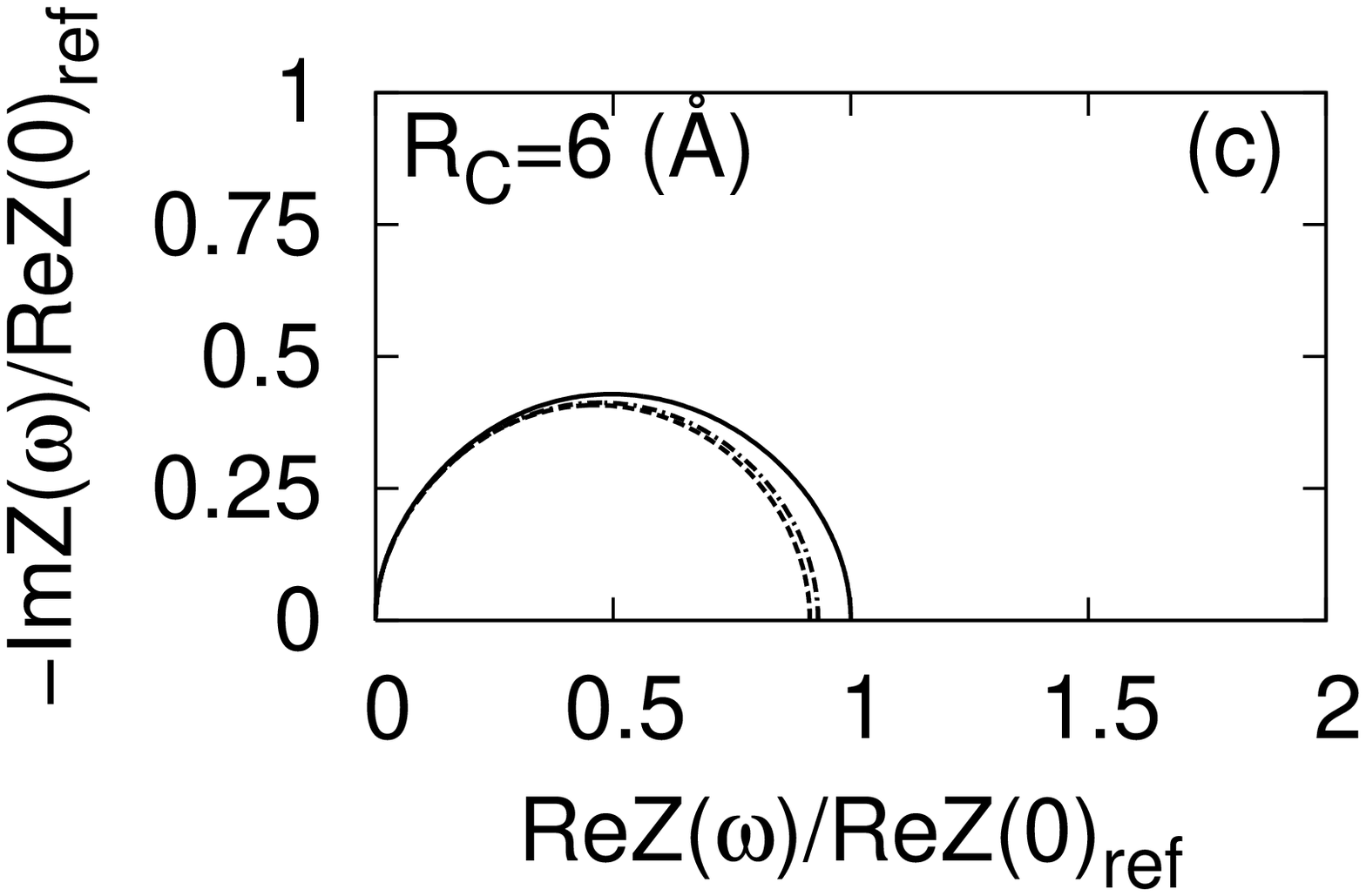} 
%\caption{minipage} 
\end{minipage} 
%\hspace{3cm}
\vspace{1cm}
        \begin{minipage}[c]{18pc}
        \includegraphics[width=0.9\textwidth]{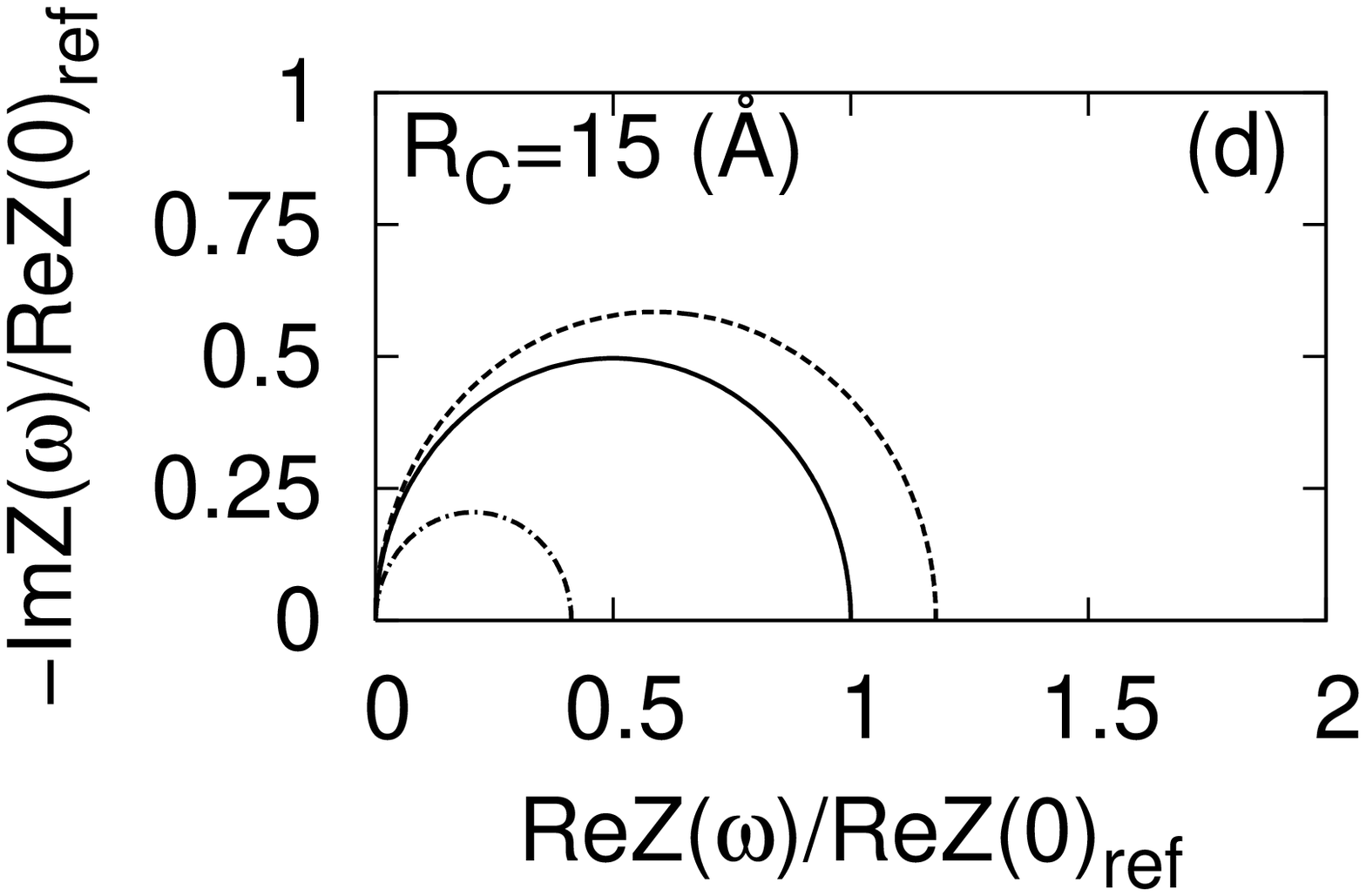} 
%\caption{minipage} 
\end{minipage} 
\caption{Nyquist plots for the native states of pR-model 1 (continuous line) and bR (dashed line), with $R_c$ = 6 \AA (figure (a)) and with $R_c$ = 15 \AA (figure (b)). Nyquist plots for the native state of pR: model 1 (continuous line), model 15 (dot-dashed line) model 4 (dotted line), with $R_c$ = 6 \AA (figure (c)) and $R_c$ = 15 \AA (figure (d)) 
} 
\label{fig:nyquist}
\end{figure}

Nyquist plot is a standard tool to display the small-signal electrical response  of a given protein \cite{Nano,Jap}.
Accordingly, \Fref{fig:nyquist} reports the theoretical  Nyquist plots calculated for the native state of  pR and bR single protein at different $R_{c}$ values. 
%
%figure13a-d
%fig7
%
Here, the shape of the Nyquist plot is found to well resemble that of a single parallel RC circuit also for small  $R_{c}$ values. 
We can observe, that the differences among the pR models are quite small at low values of the interaction radius, although they become very large at high values of  $R_{c}$. 
Furthermore, at low  $R_{c}$ values, the Nyquist plots of bR appear very different with respect to those of pR,  while at increasing $R_{c}$ values this difference becomes less significant and rather comparable with differences among different pR models. 
We conclude, that the comparison between  pR and bR models exhibits the strongest  contrast at low $R_{c}$ values. 
\subsection{Static characteristics}
In a previous paper \cite{PRE} we assumed that the \textit{bona fide} native conductance of a single protein of bR was the value deduced from AFM experiments \cite{Gomila}.
Therefore,  the INPA model parameters were calibrated to reproduce the experimental data. 
In particular, we choose the maximal/minimal resistivity as: $\rho_{max}=4\times 10^{13}\ \Omega $ \AA, $\rho_{min}=4\times 10^{5}\ \Omega$ \AA, and the barrier height associated with sequential tunneling processes between amino acids, $\Phi=$219 meV.  
Owing to the similarities between pR and bR, the same values are used for pR.
\par
\Tref{tab:1} reports the single-protein conductance, $g$, both for pR and bR, as calculated within the INPA model.
In the same Table, the conductivity $\sigma$ associated with  a 5 nm-side cube of conductance $g$, is  also reported.  
\begin{table}[htbp]
        \centering
                \begin{tabular}{l|l|l|r}
                \hline
                \hline
                & & &\\
                        \textit{protein}&\textit{state}& $g$(pS) &$\sigma$(S/cm)\\
                        && &\\
                        \hline 
                         & & \\
                        bR&dark& 0.24 & 4.8$\times 10^{-7}$\\
                        \hline 
                        & & &\\
                        bR&light&0.27&5.4$\times 10^{-7}$\\     
                        \hline 
                        & &  &\\
                        pR&dark&0.08 &1.6$\times 10^{-7}$\\
                        \hline
                        \hline
                \end{tabular}
                \caption{Theoretical values of bR an pR single-protein conductance, $g$, and conductivity, $\sigma$,  as predicted by the INPA model. }
        \label{tab:1}
\end{table}
We used the PDB files \cite{PDB} 2NTU, 2NTW and 2L6X for calculating the conductivities for native/active states of bR and  native state of pR. 
The conductivity value of pR is estimated as the mean value over the 20 structures provided by the PDB entry 2L6X. 
Accordingly, due to the different tertiary structure, the Ohmic conductivity of native bR is found to be higher than that of the corresponding pR for about a factor of 3.
The values above can be compared with the experimental values obtained from the literature as reported in \Tref{tab:2}.
\begin{table}[htbp]
        \centering
                \begin{tabular}{r|l|l|c|c}
                \hline
                \hline
                & & &\\
                        OD & \textit{protein}&\textit{state}& $g$(pS) & $\sigma$(S/cm) \\
                        & & &\\
                        \hline 
                        & &  & \\
                 10& pR & dark& 230 & $1.92\times 10^{-7}$ \\
                        \hline 
                        & & & \\
                        10& pR &light& 336 & $2.8 \times 10^{-7}$\\            
                        \hline
                         &  & &  \\ 
1& pR & dark& 105 & $8.74 \times10^{-8}$ \\
                        \hline
                         & & & \\ 
                        1& pR &light& 176 & $1.47\times 10^{-7}$ \\     
                        \hline
                        \hline
                \end{tabular}
                \caption{Conductance of pR, $g$, and conductivity, $\sigma$, as inferred from thin film experiments by Ref.~\cite{Lee}.} 
        \label{tab:2}
\end{table}
Data reported in \Tref{tab:2} are obtained after subtracting the buffer conductance, see \Fref{fig:ivst}.
We find that the pR conductivity values inferred from experiments on thin films in \Tref{tab:2} are in good agreement with the INPA values reported in Table 1. 
We also notice that, owing to the different sizes of the considered macroscopic samples, the pR conductance inferred from thin films is of about a factor $10^3$ higher than that of the single pR value in Table 1, as expected.  
%As a further consideration, we remark that the  photocurrent effect (as %the increase of current due to light) in pR is of about  68 \% at a concentration %of 1 OD and decreases to about 50 \% at 10 OD.  
%Such a decrease  should be explained with some shielding of the incomming %photons due to the growing of the superimposed layers of pR proteins.
%Notice that Fig.~2 also indicates a very sharp increase of photocurrent %at concentrations lower than 1 OD.
\par
\begin{figure}[htbp]
\centering
\includegraphics[width=0.6\textwidth]{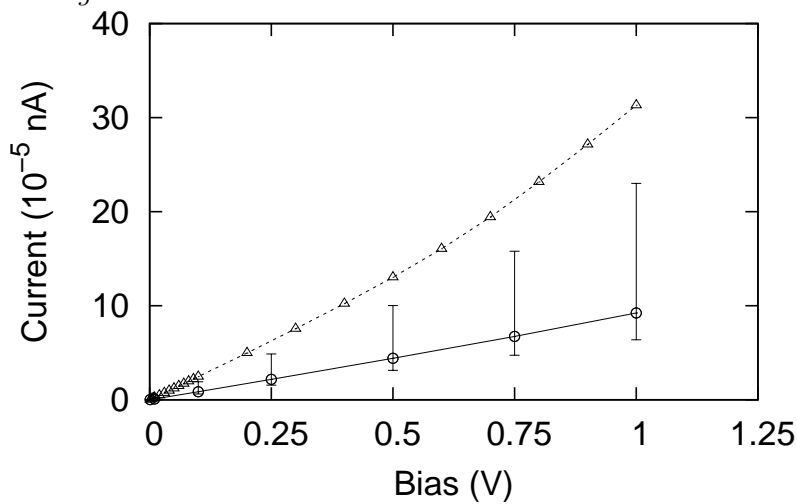}
\caption{Theoretical I-V characteristic of a single protein in its native state including sequential  tunnelling mechanism of charge transfer. Continuous line refers to pR, dashed line to bR.}
\label{fig:I-V}
\end{figure}

\Fref{fig:I-V} reports the I-V characteristics of bR and pR single proteins in their native state.
Data are calculated within the INPA model and include tunneling mechanism for charge transfer (\Eref{eq:1});  in particular, each point of the pR curve represents the mean value carried out over the available  20 models. It is evident that in the chosen bias range (0-1V) the electrical current has a quasi-Ohmic behavior, for both the proteins, i.e. the assigned barrier height is sufficiently high to depress the tunneling mechanism.

%fig8
%
\par
\begin{figure}[htbp]
\centering
\includegraphics[width=0.6\textwidth]{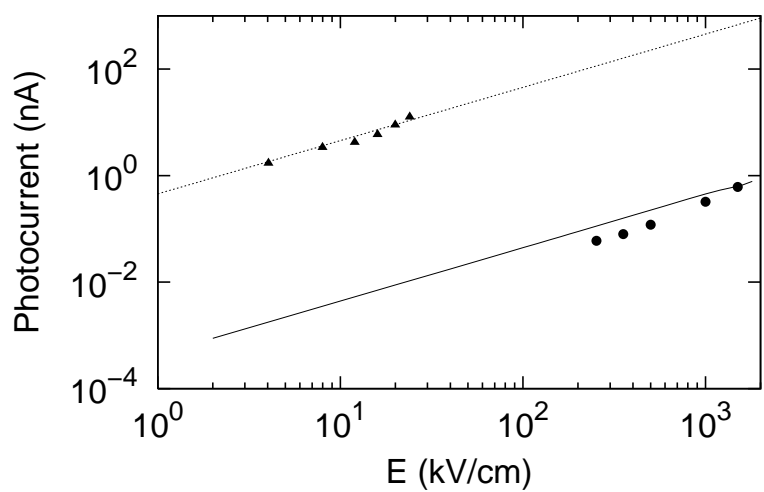}
\caption{Photocurrent as a function of the average applied electric field. 
Full circles and triangles represent experimental results  for bR
\cite{Jin} and pR \cite{Lee}, respectively. 
Continuous line refers to results of simulations carried out within the INPA model, dashed line refers to a linear fitting of pR experiments.}
\label{fig:Photocurrent}
\end{figure}
\begin{figure}[htbp]
\centering
\includegraphics[width=0.6\textwidth]{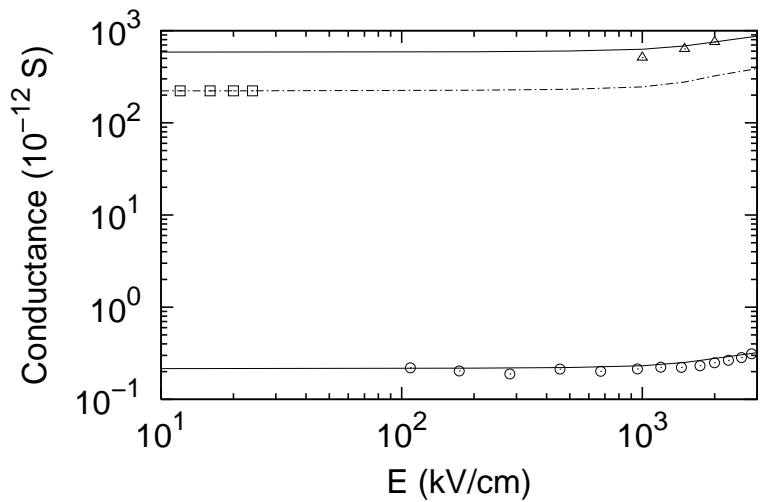}
\caption {Macroscopic protein conductance under dark conditions as a function of the applied electric field.
Symbols  refer to macroscopic experimental results, circles for \cite{Gomila}, squares for \cite{Lee}, triangles for \cite{Jin}, lines to simulations carried out by INPA modelling for a single protein.
Calculated results are fitted to data obtained at 1 V \cite{Jin,Gomila}, and to data at the lowest fields \cite{Lee}.}
\label{fig:Conductance}
\end{figure}

\Fref{fig:Photocurrent} reports the  measured photocurrent as a function of the average applied electric field,  calculated as the ratio between the applied voltage and an active length of 5 nm for the case of bR and of $5 \times 10^{-3}$ cm for pR.
The continuous curve refers to the results obtained by the INPA modelling on a single protein of bR, the dashed curve is a linear interpolation of experiments carried out on a thin film of pR.
The quasi-linear increase of the photocurrent with applied field is due to the quasi-Ohmic behaviour of the I-V characteristics.
%fig9
%
To investigate the deviation from the linear Ohmic response at increasing applied fields, \Fref{fig:Conductance} reports the conductance of pR and bR in their native states, as a function of the average applied electric-field.  
Symbols refer to the light-induced responses available from the literature \cite{Jin,Lee} and dashed curves to the results obtained by the INPA single-protein approach. 
Specifically, data by Refs.~\cite{Jin} and \cite{Lee} are fitted by changing the maximal resistivity from $4.0\times 10^{13}$ to $1.6\times 10^{10}$ in the former case and from $4.0\times 10^{13}$ to $1.5\times 10^{11}$ in the latter case.
This to take into account the macroscopic size of the samples.
\par 
The comparison between experiments and theory is forced to agree at low fields for the case of pR, and at the maximum field for bR.
For both proteins a deviation from the Ohmic behaviour  associated with a transport driven by a field dependent tunneling mechanism, is predicted at electric fields above about 1 MV/cm.
This prediction is in good agreement with experiments carried out on bR samples \cite{Jin,Gomila,PRE,PRE1}.
%fig10
%
%
\newpage
\section{Conclusions}
This paper investigates  the electrical properties of two light receptors, bacteriorhodopsin and proteorhodopsin, both pertaining to the family of $type-1$ opsins and  whose functioning is based on a proton pumping mechanism.
In particular, the photo-sensitive properties of pR were detected  using  
metal-insulator-metal thin-film structures.
The electrical changes due to different pR concentrations, different intensity and/or wavelengths of the irradiating light, were observed by measuring the real-time photocurrent characteristic of the system. 
The result of these measurements  are then compared with analogous outcomes on bR which can be found in literature.
To this purpose, an impedance protein network analogue (INPA) is used to carry out a comparative microscopic study of the 3D structures and of the conductive properties as a function of an external applied voltage.  
We assume that the charge transfer between neighbouring amino acids is the
microscopic transport mechanism which determines the macroscopic electrical properties of samples activated by these proteins. 
In particular, we suggest that a sequential tunneling mechanism is mostly responsible of the measured current - voltage characteristics at high fields.
\par
As matter of fact, by shedding on the protein sample some visible light, a photocurrent arises with a maximum centered on the green region of the spectrum.
For the case of bR, the presence of a net photocurrent was associated with the conformational change of the single protein due to the presence of light.
Because of their similarity, we expect that the analogous results reported by experiments on pR should be explained within the same mechanism. 
The results of calculations confirmed our expectation for the native state of pR.
However, the lack of knowledge of the 3D structure of pR in its active state does not allow to carry out a quantitative microscopic study of its photo-conductance.
Like experimentally found for bR, and successfully described within the INPA model, the single protein conductance of pR was determined and the small-signal response analyzed in terms of the Nyquist plots. 
Furthermore, also pR is predicted to exhibit super-linear current-voltage characteristics for applied electric fields above about 1 MV/cm.
\par
Present findings should be of relevance for a better understanding of the basic properties of transmembrane proteins, which functioning is essential for the living of a single cell.
From an applied point of view, opsin-based biomolecular electronics is a valuable premise for the development of a new generation of biodevices. 
In particular, we showed that, like bR, pR  represents a relatively simple and stable biological system for exploring electronic transport, since both proteins pertain to a biological material that can be manipulated with all the tools known to modern biophysics.
The reported experiments clearly show the sensitivity and usefulness of MIM technique when used on photo-receptors and open a new scenario for future applications in the wide field of biosensors.
%

%\begin{acknowledgments}
%
\ackn
Dr. Adriano Cola (Lecce CNR-ISAC laboratory) is thanked for useful discussions.
This research is supported by the European Commission under the Bioelectronic Olfactory Neuron Device (BOND) project within the grant agreement number 228685-2.
%\end{acknowledgments}
%
\appendix
\setcounter{section}{1}
\section*{Appendix A. The role of the interaction radius} 
%
%\section{}
%
On the side of the protein structures, for the case of bR a large number of refined models have been proposed both for native and active state.
By contrast, at present, for pR we only found one certified set of structures,  describing the protein in its native state. 
This set, taken with an  NMR technique and corresponding to the 2L6X entry of the protein data base \cite{PDB,Reckel}, consists of 20 different models  of the protein, all in principle equally able to describe the protein electrical properties.
In the following, we investigate the protein modifications by means of different tools, such as he determination of the link distributions and  the calculation of the single protein global resistance.
\begin{figure}[htbp]
\centering
\includegraphics[width=0.6\textwidth]{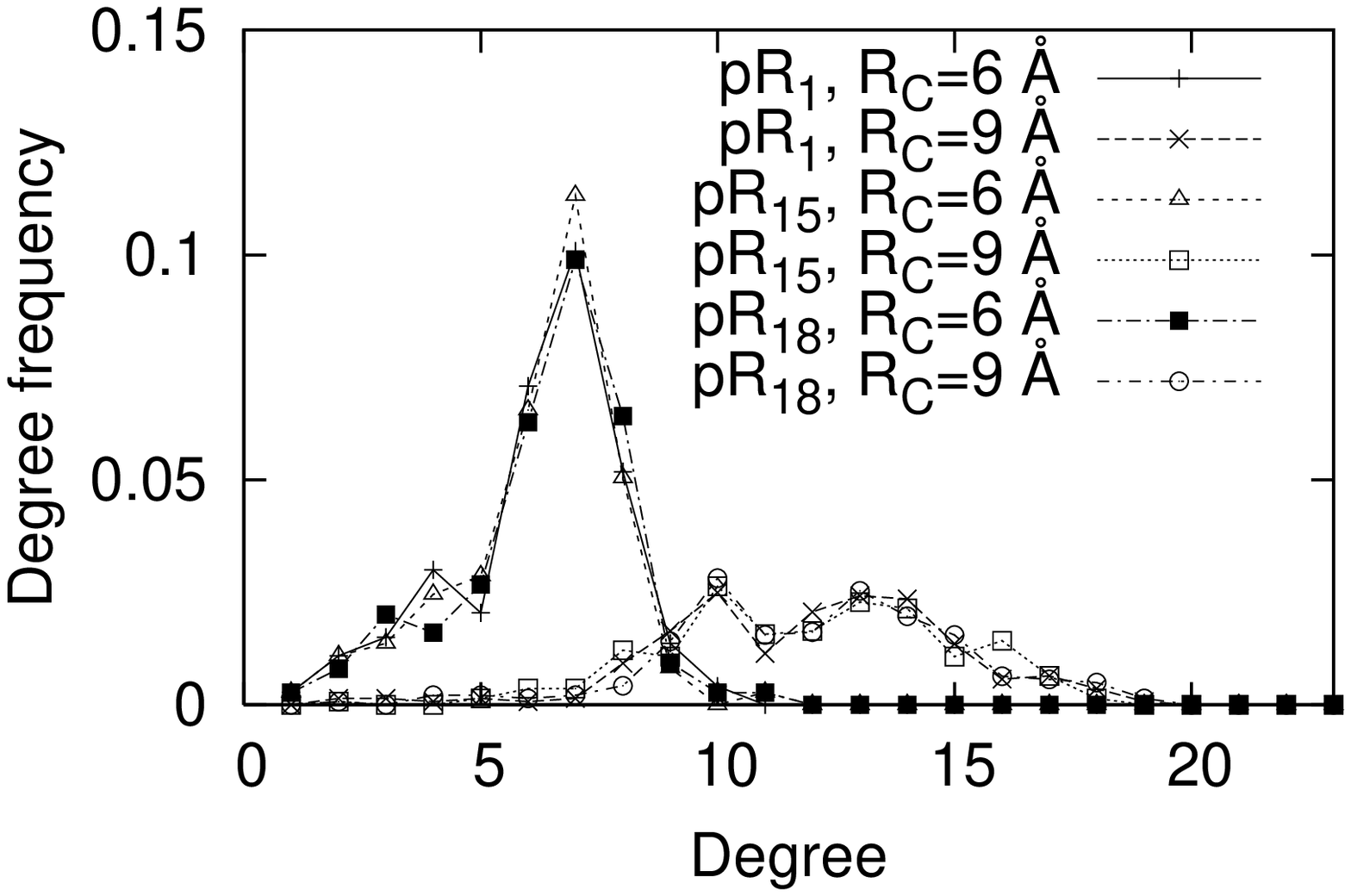}
\caption{Degree distributions for the native state of different pR models
(1,15,18) and $R_{c}=6, 9$ \AA.}
\label{fig:degree_2L6X}
%\vskip3pc
\end{figure}
%fig11
%
\par
\begin{figure}[htbp]
        \centering
                \includegraphics[width=0.6\textwidth]{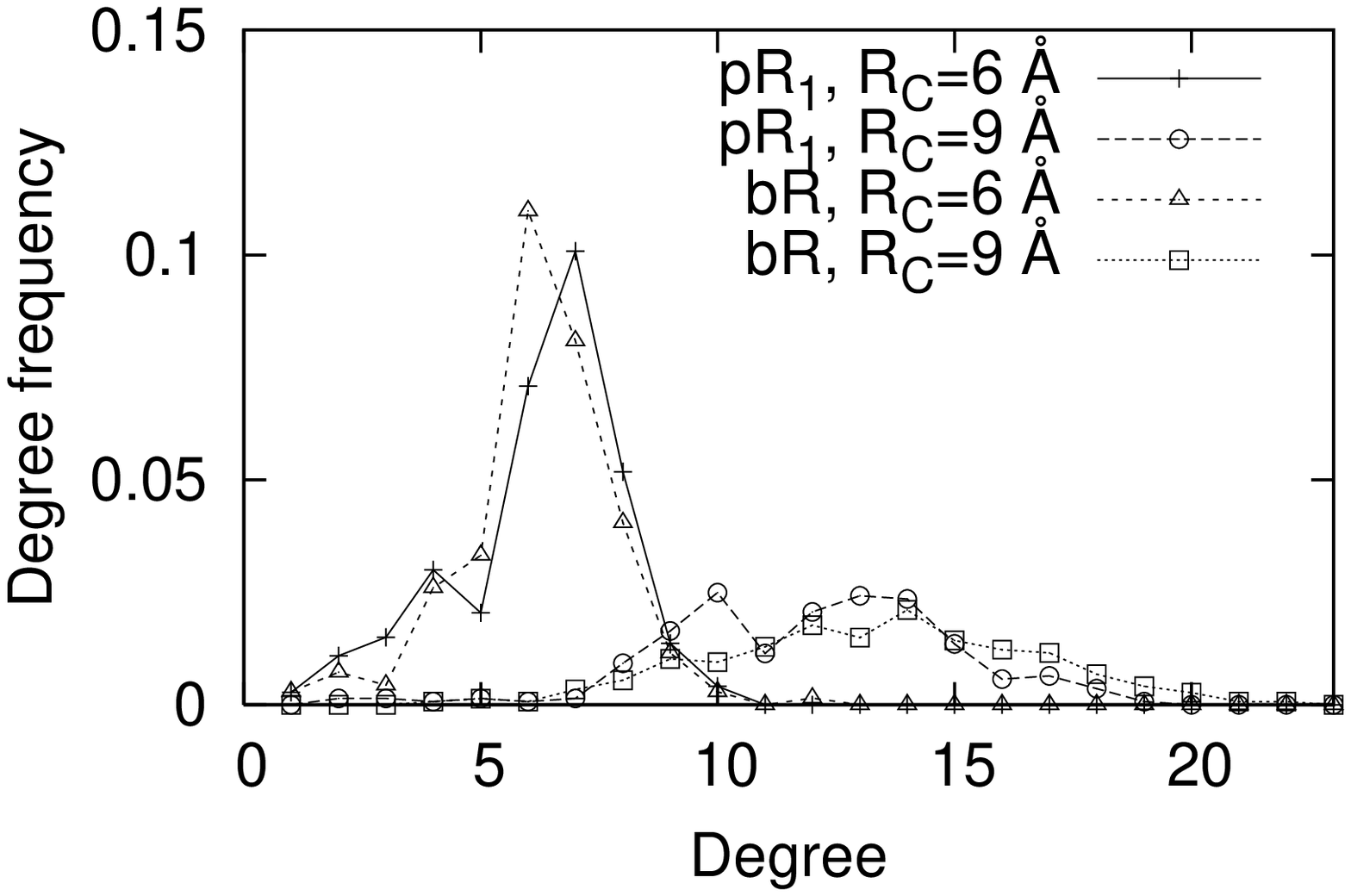}
        \caption{Degree distribution for the native state of  pR (model 1) and bR, for different $R_{c}$ values.}
        \label{fig:degree_mod1}
\end{figure}

Results  concerning the degree distribution, i.e. the number of links for each amino acid, at different $R_{c}$ values, are reported in \Fref{fig:degree_2L6X} and \Fref{fig:degree_mod1}.
In particular, \Fref{fig:degree_2L6X} compares the degree distributions of three different pR models at two values of $R_{c}$. For all the analyzed models the maximum degree is 6 at $R_{c}= 6$ \AA. 
%\vskip3pc
%fig12
%
In other terms,  independently of their position in the protein, most of amino acids has the same number (6) of nearest neighbours, a unexpected regularity. 
Furthermore, the distribution rapidly smooths down at increasing $R_{c}$  values.
\Fref{fig:degree_mod1} reports the comparison between the degree distributions of the model 1 of pR and bR. 
As noticed for the contact maps, also in this case the structures appear quite similar, although some relevant differences stand out. 
\begin{figure}[htbp]
        \centering
                \includegraphics[width=0.6\textwidth]{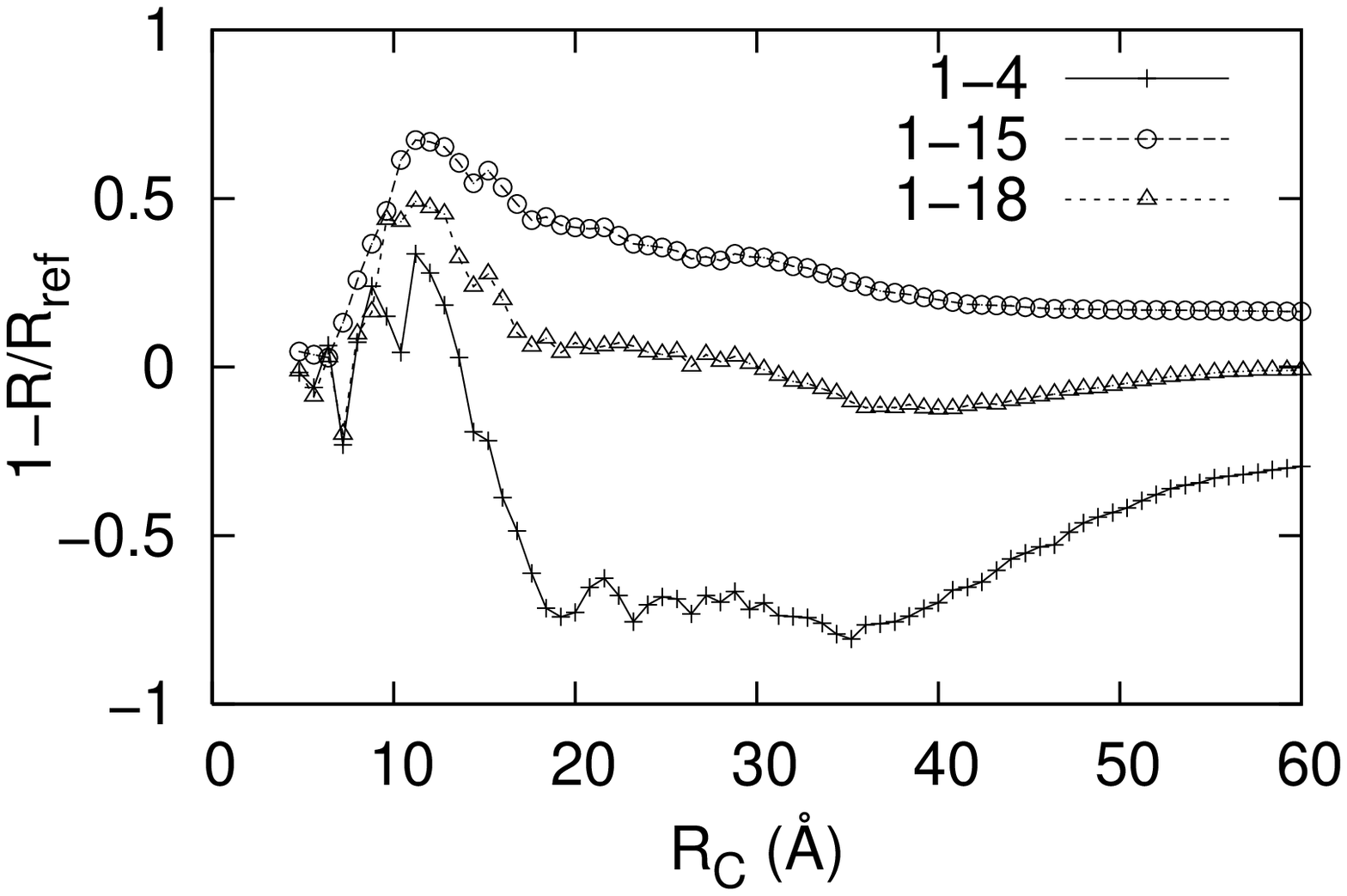}
        \caption{Relative resistance variation for the native state of the reported  pR models, at increasing values of $R_{c}$.  Curves are guides to the eye and the reference structure is the model 1 of the PDB entry.}
        \label{fig:rel_res_tot}
\end{figure}
%fig13
%
\par
By construction, the INPA model produces different electrical responses for each of the 20 NMR models. 
Therefore, it is useful to analyze these differences in order to state the boundaries of theoretical expectations on the basis of the present information.
As a first step, we analyze the single-protein resistance of the twenty models of pR, the 2L6X entry of the protein data base \cite{PDB}. 
 Some results are reported in \Fref{fig:rel_res_tot}.
\begin{figure}[htbp]
        \centering
                \includegraphics[width=0.6\textwidth]{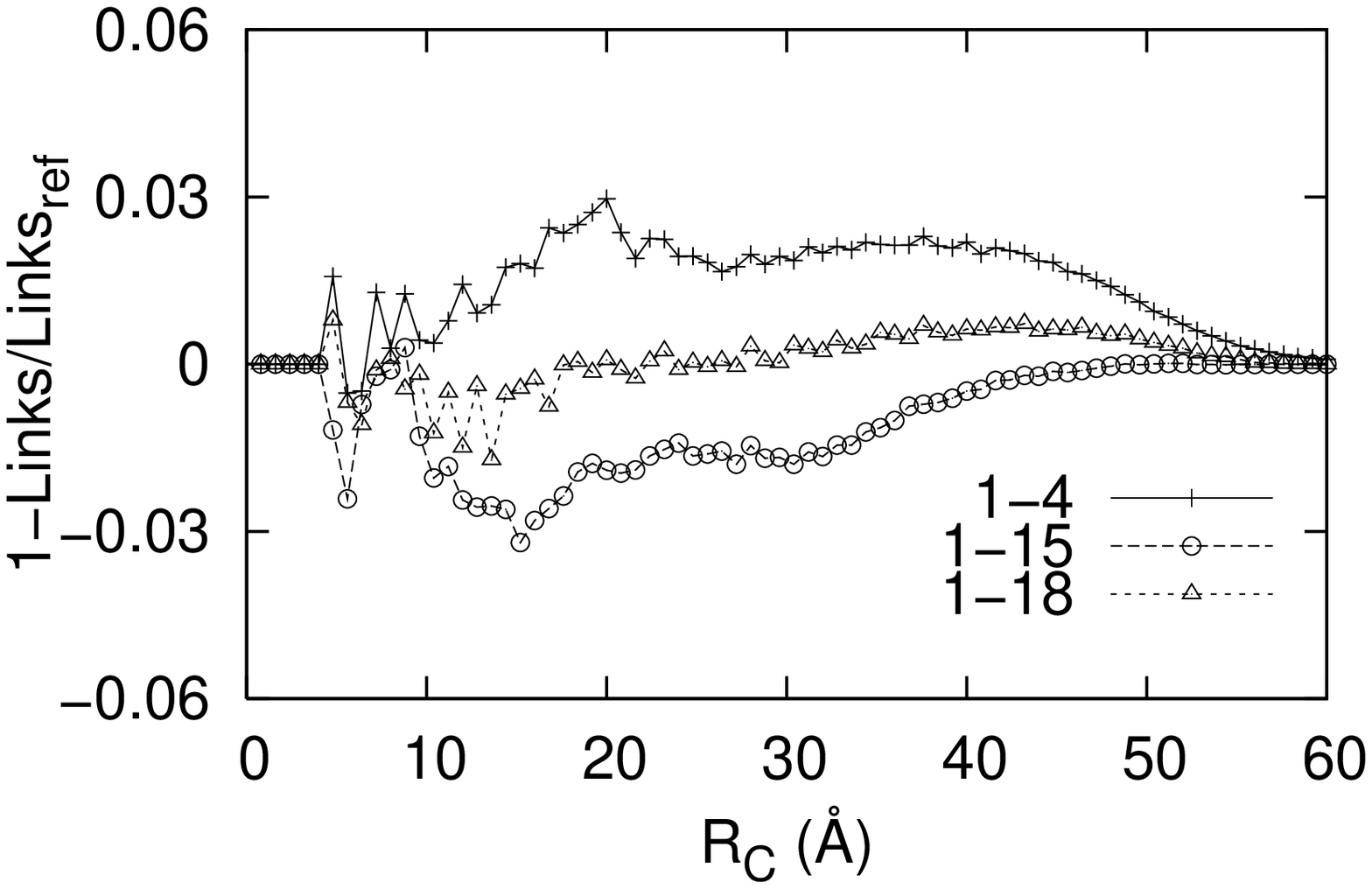}
        \caption{Relative link number for the native state of the pR models, at increasing values of $R_{c}$. Curves are guides to the eye and the reference structure is the model 1 of the PDB entry.}
        \label{fig:rel_links}
\end{figure}
%fig.14
%
Here we find that there is a peak of difference in the range of interaction radius 
$6 \div  20$ \AA. 
In particular, large deviations are found for values of  $R_{c}\sim 10$ \AA.
We notice that, while at low $R_{c}$ values the resistance of model 1 is larger than most of the other models, at  larger values of $R_{c}$ resistances follow three different pathways.
\begin{figure}
        \centering
                \includegraphics[width=0.6\textwidth]{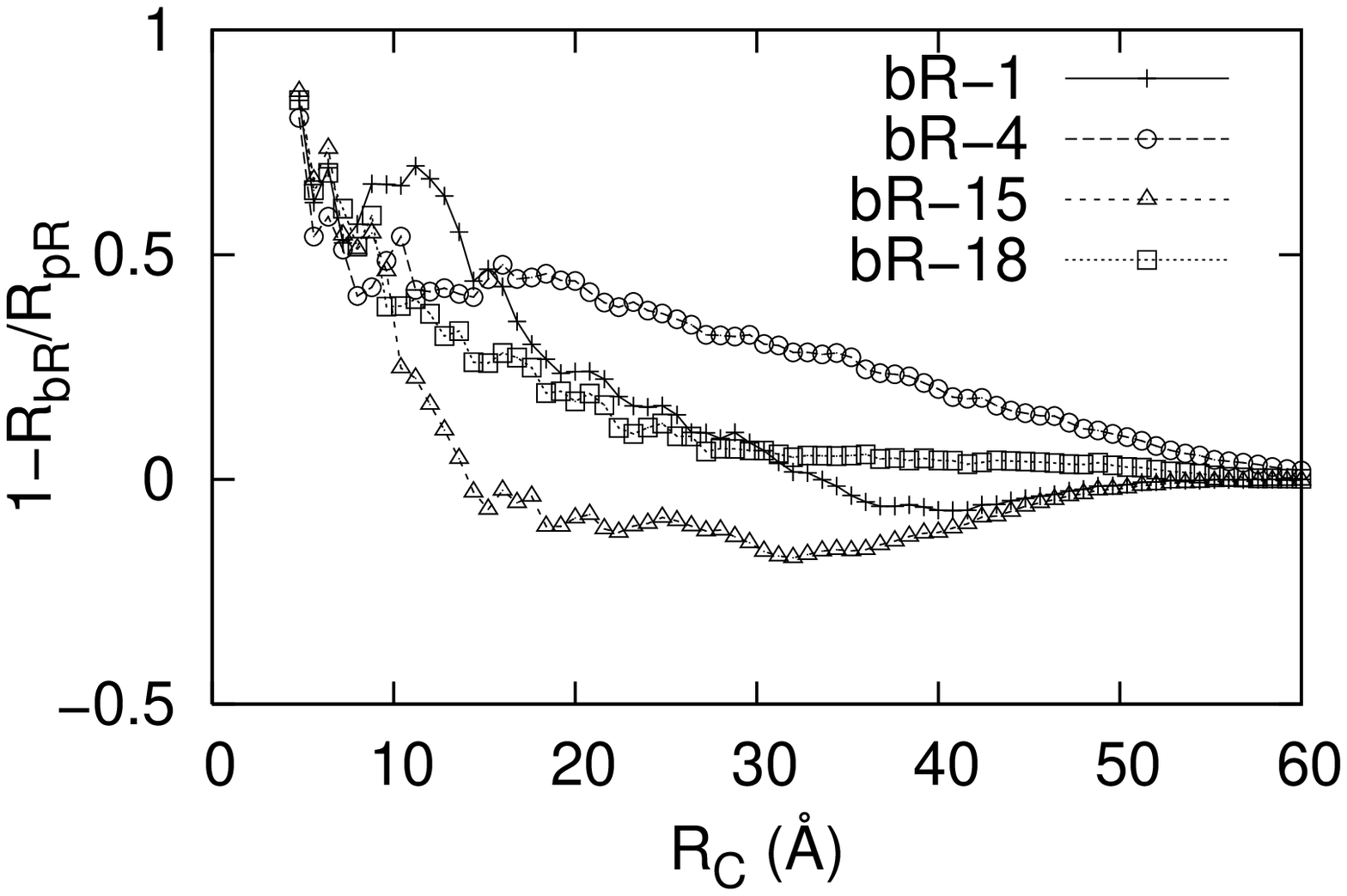}
        \caption{Relative resistance variation for the native state of the reported pR models (1,4,15,18), at increasing values of $R_{c}$.  Curves are guides to the eye and the reference structure is the native state of bacteriorhodopsin, the entry 2NTU of the PDB\cite{PDB}.}
        \label{fig:rel_res_rho}
\end{figure}
%fig15
%
(i) They can be close to that of model 1  ( e.g. model 18 in \Fref{fig:rel_res_tot}); 
(ii) they can be much greater than that of model 1 (e.g. model 4); 
(iii) they can be smaller than that of model 1  all over the $R_{c}$ range (e.g. model 15).
These peculiarities can be verified by analyzing the relative variation of the link number, as reported in \Fref{fig:rel_links}.
%\subsection{Degree distributions}
This outcome suggests the following description of the protein structures: (i) models that  preserve the relative resistance  behaviour (with respect to model 1) exhibit a helix distribution similar, and more compact, to that of model 1; 
(ii) models that change the relative resistance response exhibit a deformed helix distribution, the larger the higher $R_c$ values the smaller the lower $R_{c}$ values. 
\par
Finally, \Fref{fig:rel_res_rho} reports the relative resistance variation between the bR model 2NTU \cite{PDB} and some pR models (1,4,15,18). 
In order to take into account the different protein sizes, the comparison is performed by reporting the normalized resistance values.
Calculations show a smaller resistance of bR for nearly all the models, especially for $R_{c}$ values lower than 30 \AA, i.e. in the sensitive range. 

%
\end{document}